\documentclass[letterpaper,twocolumn]{article}

\usepackage[T1]{fontenc}

\usepackage{geometry}
\usepackage{setspace}
\usepackage{comment}
\usepackage{xcolor}

\usepackage[style = chem-acs,articletitle=true]{biblatex}
\usepackage{graphicx}
\usepackage{float}
\newfloat{scheme}{htbp}{los}
\floatname{scheme}{Scheme}
\floatname{chart}{Chart}
\newfloat{graph}{htbp}{loh}

\usepackage{chemformula} 
\usepackage[version = 4]{mhchem} 

\usepackage{hyperref}

\usepackage{authblk}
\author[1]{Samuel Del Fré}
\author[1]{Gilberto A. Alou Angulo}
\author[1]{Maurice Monnerville}
\author[1]{Alejandro Rivero Santamaría}

\affil[1]{Univ. Lille, CNRS, UMR 8523 – PhLAM – Physique des Lasers Atomes et Molécules, F-59000 Lille, France}

\title{Data-driven construction of machine-learning-based interatomic potentials for gas–surface scattering dynamics: the case of NO on graphite}

\date{*Email: samuel.del-fre@univ-lille.fr}

\begin{document}


\twocolumn[
\maketitle
\begin{center}
\begin{minipage}{0.85\textwidth}
\begin{abstract}
Accurate atomistic simulations of gas--surface scattering require potential energy surfaces that remain reliable over broad configurational and energetic ranges while retaining the efficiency needed for extensive trajectory sampling. Here, we develop a data-driven workflow for constructing a machine-learning interatomic potential (MLIP) tailored to gas--surface scattering dynamics, using nitric oxide (NO) scattering from highly oriented pyrolytic graphite (HOPG) as a benchmark system. Starting from an initial \textit{ab initio} molecular dynamics (AIMD) dataset, local atomic environments are described by SOAP descriptors and analyzed in a reduced feature space obtained through principal component analysis. Farthest point sampling is then used to build a compact training set, and the resulting Deep Potential model is refined through a query-by-committee active-learning strategy using additional configurations extracted from molecular dynamics simulations over extended ranges of incident energies and surface temperatures. 
The final MLIP reproduces reference energies and forces with high fidelity and enables large-scale molecular dynamics simulations of NO scattering from graphite at a computational cost far below that of AIMD. The simulations provide detailed insight into adsorption energetics, trapping versus direct scattering probabilities, translational energy loss, angular distributions, and rotational excitation. Overall, the results reproduce the main experimental trends and demonstrate that descriptor-guided sampling combined with active learning offers an efficient and transferable strategy for constructing MLIPs for gas--surface interactions.
\end{abstract}
\end{minipage}
\end{center}
]

\section{Introduction}
\label{sec:intro}

Gas–surface scattering plays a central role in the microscopic understanding of energy and momentum exchange at solid–gas interfaces, with direct relevance to heterogeneous catalysis, atmospheric, astrochemistry and surface science \cite{auerbachChemicalDynamicsGasphase2021,kolbOverviewCurrentIssues2010,somorjaiImpactSurfaceChemistry2011,vandishoeckAstrochemistryDustIce2014}. When an incident molecule interacts with a solid surface, energy redistribution arises from the coupled dynamics of molecular degrees of freedom and collective substrate motion \cite{arumainayagamMolecularBeamStudies1991a,diezmuinoDynamicsGasSurfaceInteractions2013, bortolaniInteractionAtomsMolecules1990}. Although experimental studies have provided detailed insight into state resolved gas surface scattering dynamics for several diatomic molecules, the theoretical studies of these processes remain necessary in order to gain a detailed understanding of the molecular-scale processes involved. These gas surface simulations remain challenging due to the high dimensionality of the problem and the need to treat both molecular motion and surface phonons with sufficient accuracy while achieving adequate statistical sampling through large ensembles of simulations \cite{diazTheoreticalSimulationsReactive2014}. The interpretation of experiments requires statistically converged, state and angle resolved observables, such as scattering angular distributions, final translational energies, and rotational or vibrational state populations. Reproducing such fine observables at the atomistic level using molecular dynamics (MD) critically depends on the underlying potential energy surface (PES), which governs the incident trajectory, interaction time, and dissipation pathways, and must remain accurate over wide ranges of incidence energies and surface temperatures to allow extensive molecular dynamics sampling.

Traditionally, gas--surface PESs are constructed using analytical functional forms or parametrized interaction models fitted to \textit{ab initio} reference data, enabling extensive classical molecular dynamics simulations (e.g., \cite{majumderChemicalDynamicsSimulation2018, greenwoodMolecularDynamicsSimulations2022, greenwoodNitricOxideScattering2021, rutiglianoScatteringDiatomicMolecules2024}). However, capturing the full complexity of gas–surface interactions over broad configurational spaces, particularly when accounting for surface thermal motion and wide ranges of incidence energies, poses increasing challenges for parametrized representations. \textit{Ab initio} molecular dynamics (AIMD) based on density functional theory (DFT) provides a more comprehensive description \cite{riverosantamariaInitioMolecularDynamics2019a,alouanguloInitioMolecularDynamics2024, omarHowSurfaceReconstruction2025, kroesInitioMolecularDynamics2014} but remains computationally prohibitive for the large trajectory ensembles required to converge scattering observables.
In this context, machine-learning interatomic potentials (MLIPs) offer a practical route to combine near-\textit{ab initio} fidelity with the efficiency needed for large-scale trajectory sampling. \cite{jiangHighFidelityPotentialEnergy2020, liuConstructingHighDimensionalNeural2018, riverosantamariaHighDimensionalAtomisticNeural2021, onatSensitivityDimensionalityAtomic2020}. 

A key challenge in constructing high-quality MLIPs is to build training datasets that adequately cover the regions of configurational space relevant to gas-surface interaction while avoiding unnecessary oversampling of already well-represented regions \cite{finkbeinerGeneratingMinimalTraining2024,zhangActiveLearningUniformly2019a}. This is particularly important for rare but dynamically relevant events such as close approaches and high-energy impacts on thermally excited surface configurations. Addressing this challenge requires controlled sampling strategies capable of identifying the most informative configurations within existing \textit{ab initio} datasets. In that sense, descriptor-based sampling approaches have been shown to improve the efficiency and robustness of MLIPs by explicitly characterizing and balancing coverage of high-dimensional feature spaces \cite{imbalzanoAutomaticSelectionAtomic2018,liAutomaticFeatureSelection2024,perezBitBIRCHEfficientClustering2025,qiRobustTrainingMachine2024a,liLocalenvironmentguidedSelectionAtomic2024}. 
In addition, the use of atomic descriptors enables an explicit visualization of the underlying chemical space, either in the original high-dimensional representation or after dimensionality reduction \cite{musilPhysicsInspiredStructuralRepresentations2021,chengMappingMaterialsMolecules2020,capelliDataDrivenDimensionalityReduction2021}, thereby providing a physically interpretable framework to assess configurational diversity and to guide systematic, coverage-driven sampling strategies.
A complementary challenge is to enable the targeted enrichment of the training dataset by selectively introducing new electronic-structure calculations only when required, as achieved through active learning strategies \cite{mikschStrategiesConstructionMachinelearning2021,smithLessMoreSampling2018a}.

In the present work, we apply this strategy of optimized configurational space sampling and active learning to build a MLIP tailored for nitric oxide (NO) scattering from highly oriented pyrolytic graphite (HOPG). NO scattering from graphite constitutes a long standing benchmark for interfacial energy transfer, with early molecular beam and laser based studies providing state resolved angular, velocity, and rotational energy distributions over a range of surface temperatures and collision conditions \cite{frenkelRotationalStatePopulations1982,hagerStateselectiveVelocityAngular1985a,hagerLaserInvestigationDynamics1985,hagerScatteringRotationallyExcited1992,}. In particular, detailed measurements have established the coexistence of quasi specular and diffuse scattering components and have quantified rotational energy transfer and partial accommodation effects \cite{vachEnergyTransferProcesses1987,kuzeInfluenceScatteringHistory1988a,vachSurvivalRelaxationExcitation1989,hagerRotationallyExcitedNO1997,hagerScatteringNOGraphite2004}. The availability of such state resolved observables, together with extensive prior modeling efforts \cite{nymanMonteCarloTrajectory1987b,petterssonClassicalTrajectoryStudy1988a,nymanSurfaceScatteringNO1990,rutiglianoScatteringNOMolecules2025}  makes NO on graphite a well suited model system for developing accurate potential energy surfaces and for gaining detailed atomistic insight into the mechanisms underlying gas surface scattering. Although direct quantitative comparisons between simulations and experiments require care due to differences in beam conditions, incidence geometries, surface preparation, and the time scales probed (picoseconds in simulations versus much longer experimental time scales), atomistic simulations nevertheless provide valuable insight into the microscopic mechanisms governing the observed trends. Beyond this fundamental interest, nitric oxide also plays a central role in atmospheric chemistry as a key component of the NO$_x$ cycle \cite{devriesImpactsNitrogenEmissions2021}, while graphite provides a widely used model surface for carbon-based materials \cite{xiReviewRecentResearch2021, sainiCarbonNanomaterialsDerived2021}.

Accordingly, starting from an initial AIMD dataset, we perform controlled sampling in a reduced, fingerprint-based representation of local atomic environments, and then iteratively refine the training set using a query-by-committee (QBC) active-learning scheme. The resulting trained Deep Potential (DP) type MLIP is validated, and subsequently employed to perform extensive MD simulations of NO scattering from graphite slab at normal incidence, covering collision energies from 0.05 to 2.0 eV at a surface temperature of 100 K, together with additional simulations at a fixed incident energy of 0.1 eV over surface temperatures of 50, 100, 300, and 500~K to probe temperature-dependent trends. While demonstrated here for the NO–graphite system, the workflow is general and can be extended to other gas–surface systems. 

The article is structured as follows. Section~\ref{sec:methods} presents the computational methodology, including the AIMD reference calculations, the descriptor-based sampling strategy used to construct the initial training set, the dimensionality-reduction analysis of configurational diversity, the MLIP training procedure, the molecular dynamics setup, and the active-learning refinement protocol. Section~\ref{sec:results_discussion} reports the results and discussion, first addressing the training and validation performance of the MLIP together with the efficiency of the active-learning strategy, and then presenting the molecular dynamics simulations of NO scattering from graphite, including adsorption energetics, scattering probabilities, translational energy loss, angular distributions, and rovibrational excitation. Finally, Section~\ref{sec:ccl} summarizes the main conclusions and perspectives.

\section{Methods}
\label{sec:methods}

The overall workflow comprises several successive steps. First, controlled sampling is performed on an existing AIMD dataset of NO–graphite scattering simulations to construct an initial training set. Second, MLIPs are trained on this dataset to form a model committee. Third, molecular dynamics simulations are carried out over a range of surface temperatures and incidence energies using one member of the committee to generate candidate configurations for active learning. Fourth, configurations identified as requiring improved description are selectively labeled through additional DFT calculations. Finally, the potential is fine-tuned by incorporating these newly labeled configurations into the training set.

\subsection{AIMD calculations} 
\label{subsec:AIMD}

The \textit{ab initio} molecular dynamics (AIMD) simulations were performed following protocols established in previous studies of NO and O$_2$ scattering from pristine and oxidized HOPG surfaces \cite{alouanguloInitioMolecularDynamics2024,riverosantamariaInitioMolecularDynamics2019a}.
Reference AIMD simulations were carried out within DFT using the Vienna \textit{Ab initio} Simulation Package (VASP) package \cite{kresseEfficientIterativeSchemes1996a,kresseEfficiencyAbinitioTotal1996a}. The Perdew--Burke--Ernzerhof (PBE) exchange–correlation functional \cite{perdewGeneralizedGradientApproximation1996a}  with D3(BJ) dispersion correction \cite{grimmeConsistentAccurateInitio2010,grimmeEffectDampingFunction2011} was employed together with projector augmented-wave (PAW) pseudopotentials \cite{blochlProjectorAugmentedwaveMethod1994a}. The graphite surface was modeled using a periodic $4\times4\times3$ slab exposing the (0001) basal plane.
After equilibration of the slab at surface temperatures of 100 and 300~K using a Nosé–Hoover thermostat, gas–surface scattering trajectories of NO were simulated in the microcanonical ensemble using quasi-classical trajectories (QCT). The NO molecule was initialized in its ground rovibrational state ($v=0$, $j=0$) with the incident velocity oriented normal to the surface. Two incidence energies, 0.1 and 0.3~eV, were considered, and 100 trajectories were propagated for each combination of surface temperature and incident energy.
Further computational details are provided in the Supporting Information.

\subsection{Sampling of initial training set}
\label{subsec:sampling}

Local atomic environments for all structures generated during the AIMD simulations were represented using smooth overlap of atomic positions (SOAP) descriptors \cite{bartokRepresentingChemicalEnvironments2013}. The descriptors were computed using the \textsc{DScribe} library \cite{himanenDScribeLibraryDescriptors2020,laaksoUpdatesDScribeLibrary2023} with a cutoff radius of 6~\AA. The SOAP expansion employed $n_{\text{max}} = 4$ radial basis functions and $l_{\text{max}} = 4$ spherical harmonics, together with $\mu_2$ power spectrum compression \cite{darbyCompressingLocalAtomic2022}, resulting in a 50-dimensional descriptor vector for each atomic environment. A Gaussian smearing width of $\sigma = 1.0$ was used. SOAP descriptors are employed here, mainly because of their demonstrated effectiveness for pattern recognition and structural analysis in atomistic configurations (e.g., \cite{leithererRobustRecognitionExploratory2021,deComparingMoleculesSolids2016,laiFuzzyClassificationFramework2023,delfreUnsupervisedMachineLearningPipeline2025}) but the sampling workflow itself is general and can be readily applied using other vector-based atomic fingerprints.
Once the SOAP descriptors were computed for all AIMD configurations, the resulting descriptor vectors were combined into a single global feature matrix. Dimensionality reduction was then performed using principal component analysis (PCA) \cite{jolliffePrincipalComponentAnalysis2016}, retaining the minimum number of principal components required to capture 95\% of the total variance of the SOAP feature space. This reduced representation was subsequently used for data selection and sampling.
Sampling of representative environments was then performed in the reduced feature space using farthest point sampling (FPS) \cite{liuFarthestPointSampling2025,imbalzanoAutomaticSelectionAtomic2018, williamsHessianQM9Quantum2025}. 
The FPS procedure was performed iteratively in batches of 100 configurations, with the coverage of the selected principal components evaluated after each batch. Sampling was continued until 99~\% of the range spanned by the retained principal components was covered, thereby defining a data-driven stopping criterion that ensures broad and balanced sampling of the reduced configurational space without imposing a predefined sample size.
FPS was applied at the level of atomic environments. When an atomic environment was selected in the reduced feature space, the corresponding full atomic configuration was retained. 

\subsection{MLIP training}
\label{subsec:mlip_train}

Machine-learning interatomic potentials were trained using the Deep Potential formalism as implemented in \textsc{DeePMD-kit} \cite{wangDeePMDkitDeepLearning2018a,zengDeePMDkitV2Software2023a,zengDeePMDkitV3MultipleBackend2025} with hyperparameters adapted from our previous study \cite{infusoDeepPotentialdrivenMolecular2025}. Local atomic environments were represented using the two-body embedding DeepPot-SE descriptor \cite{zhangEndtoendSymmetryPreserving2018} with an outer cutoff radius $r_\mathrm{cut}=5.5$~\AA\ and a smoothing length $r_\mathrm{cut,smth}=0.5$~\AA. The embedding network consisted of three hidden layers of 25, 50 and 100 neurons each, and an axis embedding dimension of 16. The fitting network used three hidden layers of 250 neurons each. Both descriptor and fitting network use with hyperbolic tangent activation function.

The potential was trained to reproduce DFT energies and forces using an exponential learning-rate schedule, decreasing from $10^{-3}$ to $3.5\times10^{-8}$ with a decay step of 5000. Training was performed for $10^{6}$ optimization steps using an automatically determined batch size (equal to 1 for the present system).
The FPS-sampled dataset (see previous subsection) was split into 80\% for training and 20\% for validation, and learning curves were monitored throughout training. To construct the committee used for active learning, four independent models were trained using different random seeds.

\subsection{Molecular dynamics setup with the MLIP}
\label{subsec:md_mlip}

Classical MD simulations were performed using the MLIP within the Large-scale Atomic/Molecular Massively Parallel Simulator (\textsc{LAMMPS}) \cite{thompsonLAMMPSFlexibleSimulation2022}, DeePMD-kit module. Prior to the production runs, the surface geometry was optimized using a conjugate gradient algorithm as implemented in \textsc{LAMMPS}. All canonical and microcanonical molecular dynamics simulations were carried out using a time step of 0.1~fs. Unless stated otherwise, MD simulations were performed using an $8\times8\times3$ graphite slab (384 atoms). This larger slab (compared to the $4\times4\times3$ slab used for the AIMD reference calculations) was employed in the production simulations to further limit any possible finite-size effects and to ensure that surface atomic displacements induced at elevated surface temperatures and high collision energies remain well accommodated within the simulation cell, without spurious interactions between periodic replicas.
Thermal equilibration at the target surface temperatures was carried out in the canonical ensemble using a Nosé–Hoover thermostat with a damping parameter of 20~fs for a total duration of 20~ps. Gas–surface scattering QCT were subsequently conducted in the microcanonical ensemble following the same standard protocol as used for the AIMD reference calculations. In each simulation, the NO molecule is initialized in its vibrational and rotational ground state $(v = 0, j = 0)$. Each scattering trajectory was propagated for a maximum simulation time of 6~ps and was terminated earlier if the center of mass of the scattered NO molecule reached a distance greater than 7~Å from the surface. The translational, vibrational, and rotational energies of the molecules along the trajectories were evaluated using the standard semiclassical energy decomposition procedure. 

\subsection{Active-learning based refinement protocol}
\label{subsec:active_learning}

MLIP model improvement through active learning was carried out using a MD exploration strategy based on a query-by-committee framework. Candidate atomic configurations were generated from classical MD simulations using one member of the four-model Deep Potential committee as the interatomic potential, which had been trained on the FPS-sampled dataset. Unless stated otherwise, all molecular dynamics parameters and scattering protocols were identical to those described in the previous subsection.

For the active-learning exploration, the $4\times4\times3$ DFT-optimized graphite slab was first re-optimized using the selected Deep Potential model and subsequently equilibrated in the canonical ensemble at surface temperatures of 50, 100, 200, 300, 400, and 500~K. Scattering trajectories were then propagated in the microcanonical ensemble with incident energies of 0.01, 0.025, 0.05, 0.1, 0.3, 0.5, 0.8, 1.0, 1.2, 1.5, 1.8, and 2.0~eV. For each combination of incident energy and surface temperature, five independent trajectories were generated, yielding a total of 360 simulations.

During the exploration runs, atomic forces were independently predicted by all four models of the committee for configurations sampled every 5~fs along the trajectories. Model uncertainty was quantified using the standard deviation of the atomic force components across the committee. A configuration was identified as informative and selected for subsequent DFT single-point calculations when the force deviation satisfied
\begin{equation}
0.05~\mathrm{eV/\text{\AA}} \le \Delta F \le 0.5~\mathrm{eV/\text{\AA}}.
\end{equation}
where $\Delta F$ denotes the maximum atomic force deviation within the configuration. Configurations with deviations below the lower threshold were considered well described by the current potential, while those exceeding the upper threshold were excluded to avoid unphysical configurations.

The selected configurations were extracted and subsequently labeled through single-point DFT calculations to obtain reference energies and forces with VASP. These newly labeled configurations were added to the sampled dataset, which was then repartitioned into training and validation subsets following an 80/20 split, and the Deep Potential models were retrained, closing the active-learning loop. The retraining step consisted of fine-tuning the existing models for an additional $3\times10^{5}$ optimization steps using an exponential learning-rate schedule, with the learning rate decreasing from $1.0\times10^{-4}$ to $1.0\times10^{-8}$ and a decay step of 5000.

\section{Results and discussion}
\label{sec:results_discussion}
\subsection{MLIP training and validation}
\label{subsec:mlip_valid}

 \begin{figure}[h]
     \centering
      \includegraphics[width=0.95\linewidth]{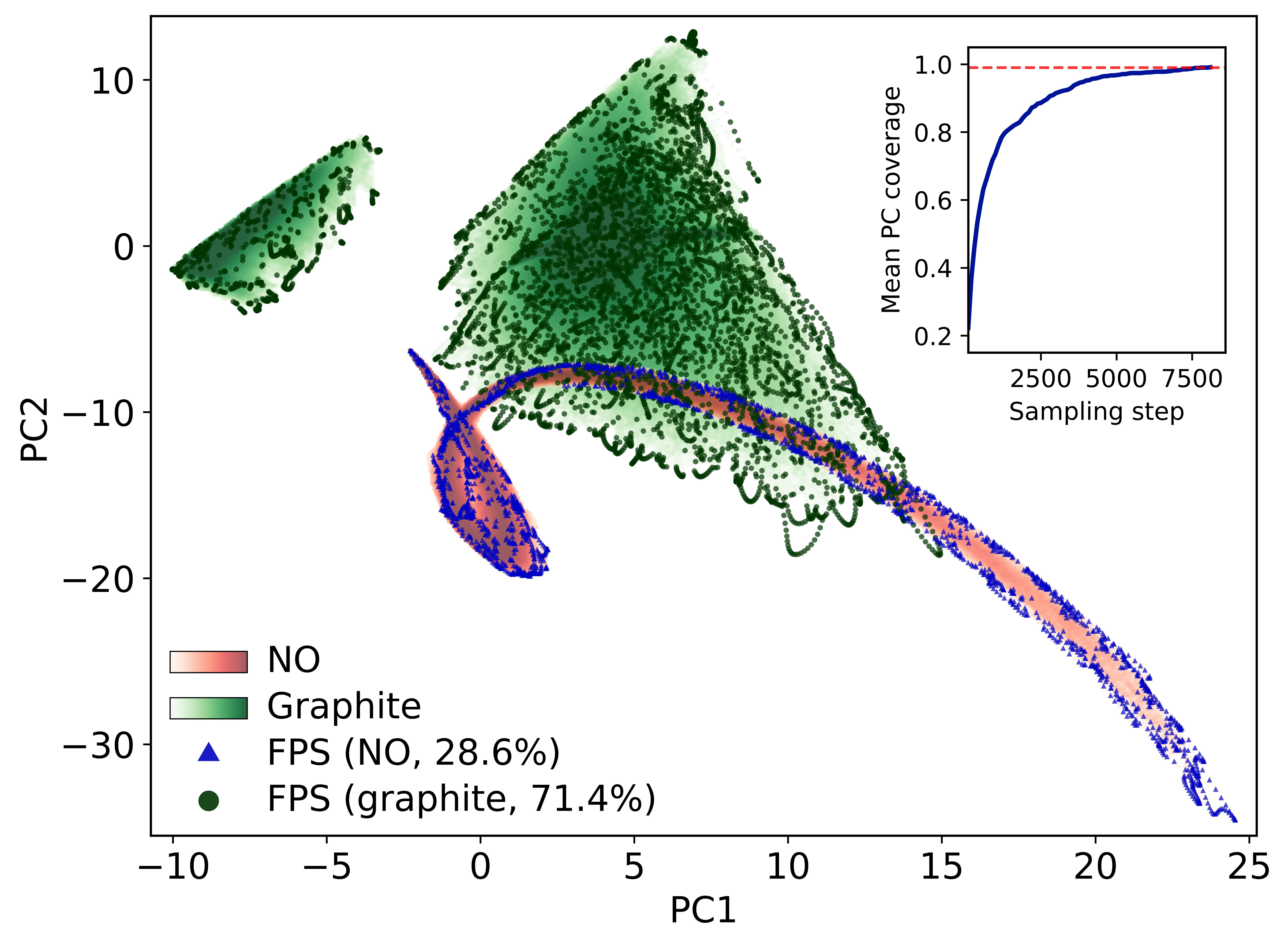}
\caption{PCA projection of the SOAP descriptor space for the full AIMD dataset, with FPS-selected environments highlighted as blue triangles (NO) and dark green dots (graphite). The inset shows the coverage evolution.}
\label{fig:Fig_b}
\end{figure}

The initial set of 400 AIMD scattering trajectories comprises a total of $N_{\text{st}} = 742 046$ DFT configurations, corresponding to more than 72 million local atomic environments when considering all atoms in each structure. This motivates the use of dimensionality reduction and sampling procedures to construct a tractable and representative training dataset.

Figure~\ref{fig:Fig_b} illustrates the descriptor-based configurational space of the initial AIMD dataset, visualized in the space spanned by the first two principal components (PCs) of the SOAP feature space. Atomic environments selected by FPS algorithm are overlaid on this representation, highlighting how the sampling procedure spans the full extent of the reduced configurational space.
First, dimensionality reduction reveals that the SOAP representation is highly compressible: the first four principal components account for 95\% of the total variance, with individual contributions of approximately 71\%, 11\%, 9\%, and 4\%, respectively. This indicates that the essential structural diversity of the dataset can be captured in a reduced four-dimensional feature space, effectively compressing the original 50-dimensional SOAP descriptor without significant loss of information. 
The full set of atomic environments is projected onto this reduced representation, revealing separation between environments associated with the NO molecule (red shaded region) and those associated with the graphite substrate (green shaded region). As expected, the graphite environments populate a broad, dense region of the feature space, reflecting the diversity of local surface configurations arising from thermal motion, while the NO-related environments form a more elongated and structured manifold corresponding to variations in molecular orientation, distance from the surface, and interaction strength along the scattering trajectories.  

Interestingly, FPS systematically selects environments that maximize coverage of the reduced feature space, leading to a uniform and balanced representation of both molecular and surface configurations without bias toward densely populated regions. As indicated by the sampling statistics shown in the inset of Figure~\ref{fig:Fig_b}, only 8100 FPS steps were sufficient to reach 99\% coverage of the atomic descriptor space. Among the sampled atomic environments, 71.4\% correspond to graphite atoms and 28.6\% to NO atoms. This resulted in the selection of 6671 distinct DFT configurations, corresponding to approximately 0.9\% of the full AIMD dataset, hereafter referred to as dataset~A. This substantial reduction reflects the presence of significant overlap among local atomic environments in the original trajectories and demonstrates that a compact subset can capture the essential configurational diversity relevant for MLIP training.

\begin{figure}[h]
    \centering
    \includegraphics[width=1.05\linewidth]{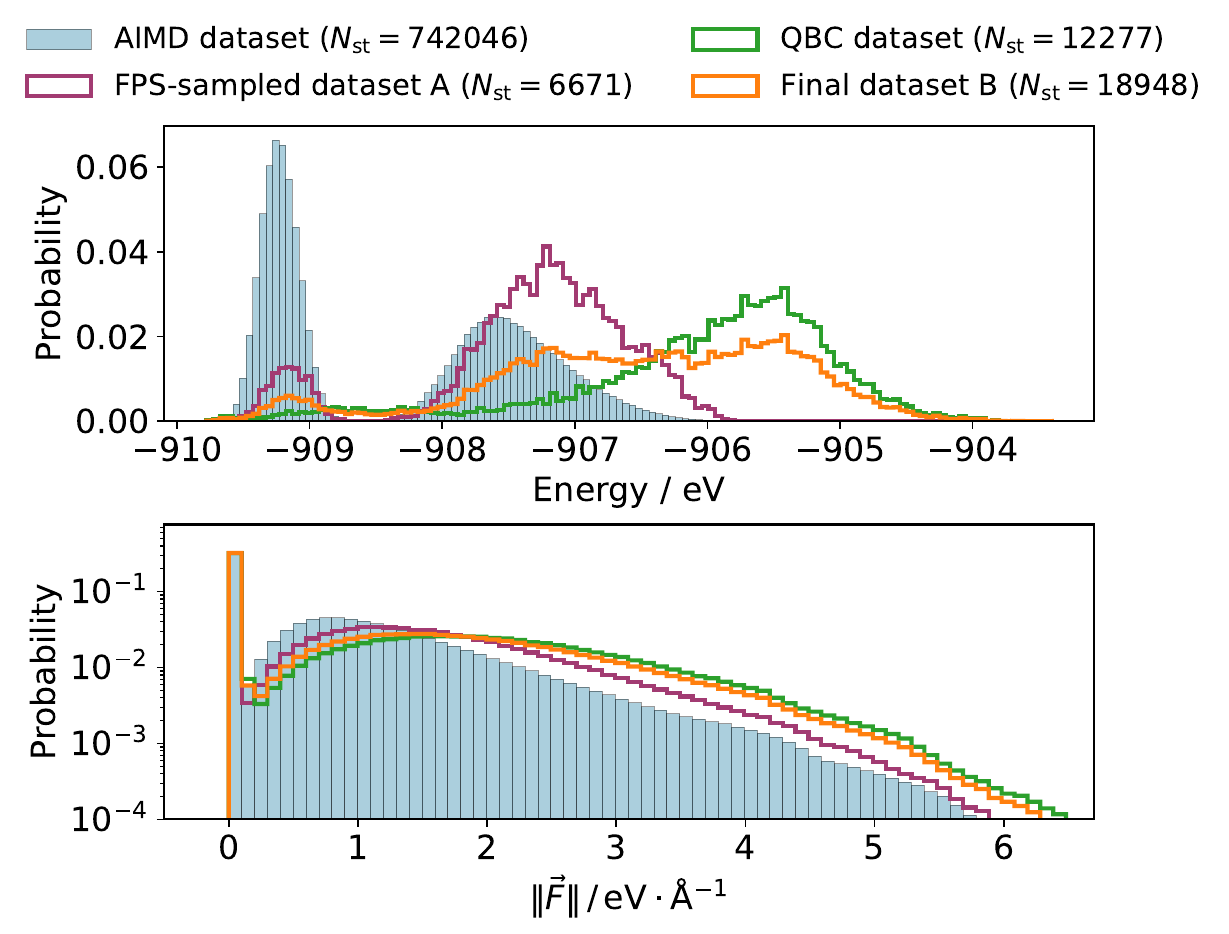}
    \caption{\textcolor{black}{Energy (upper panel) and force (lower panel) distributions for the full AIMD dataset, the FPS-sampled dataset A, the QBC-selected configurations, and the final refined dataset B. Force magnitudes are shown on a logarithmic scale.}}
    \label{fig:Fig_e}
\end{figure}

As shown in the upper panel of Figure~\ref{fig:Fig_e}, the distribution of total energies in the sampled dataset~A (purple line) closely follows that of the full AIMD dataset (blue distribution), capturing both the dominant low-energy configurations and the higher-energy tail arising from simulations performed at elevated surface temperature (300~K in this case). The sampled distribution is slightly broadened toward higher energies, reflecting the fact that FPS preferentially retains configurations located at the boundaries of the descriptor space. This behavior is intrinsic to the algorithm, which is designed to maximize coverage of the feature space and therefore tends to select less frequently visited configurations that often correspond to local energetic extrema or transition regions of the potential energy surface. Such configurations, while rare in the raw trajectories, are dynamically important for constructing a robust MLIP. A similar trend is observed for the distribution of DFT force magnitudes, represented in the lower panel of Figure~\ref{fig:Fig_e}. The sampled dataset reproduces the overall shape of the force distribution of the full AIMD dataset over several orders of magnitude, while exhibiting a slight shift toward larger force values, consistent with the preferential inclusion of configurations located in high-force regions of the configurational space. Importantly, no gaps or missing regions are observed in either the energy or force distributions, indicating that the sampling procedure preserves continuity of the underlying configurational space spanned by the initial AIMD trajectories. This continuity is essential to avoid extrapolation artifacts when training the MLIP on the sampled subset. At the same time, it should be noted that the energy distributions of the full AIMD dataset exhibit distinct subpopulations associated with the different simulation conditions employed, most notably the presence of two separated energy distributions corresponding to the two surface temperatures considered. This leads to regions of the potential energy surface that are not sampled in the initial AIMD data. As a consequence, molecular dynamics simulations performed outside the original AIMD conditions would inevitably enter extrapolative regimes if no additional reference data were introduced, as expected. This observation motivates the subsequent use of active learning to iteratively extend the training dataset beyond the configurational space initially explored by AIMD, while maintaining continuity and controlled coverage of the underlying PES.

\begin{figure}[h]
    \centering
    \includegraphics[width=\linewidth]{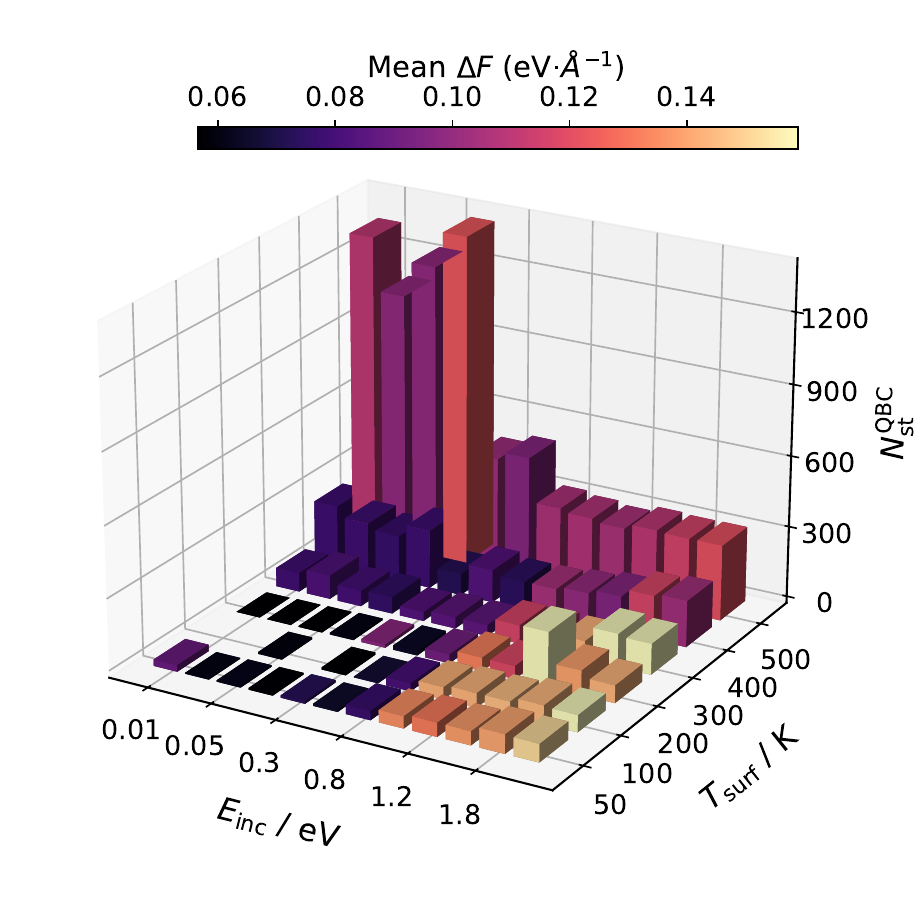}
    \includegraphics[width=\linewidth]{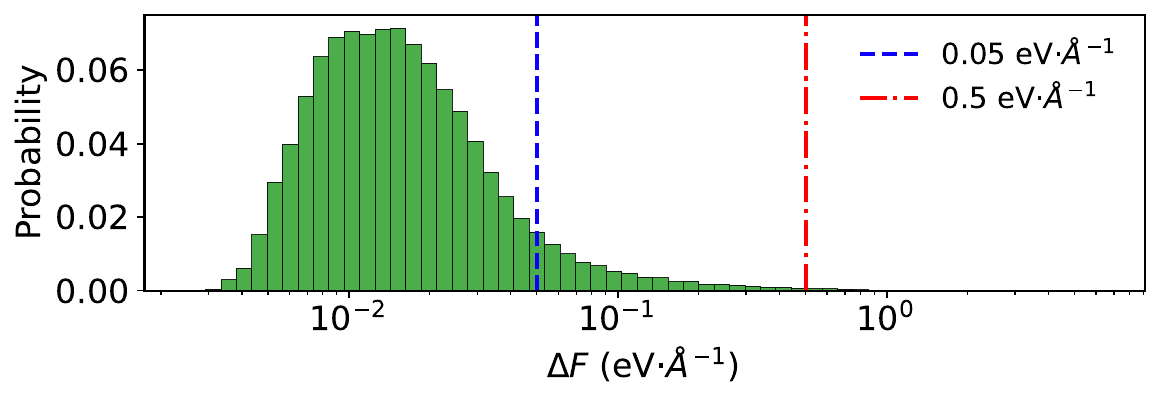}
    \caption{Three-dimensional representation of the $N_{\mathrm{st}}^{\mathrm{QBC}}$ configurations selected during the active learning phase as a function of surface temperature $T_{\mathrm{surf}}$ and incident energy $E_{\mathrm{inc}}$ (upper panel). Distribution of the QBC model-deviation score ($\Delta F$) over the extracted configurations on a logarithmic scale, with the selected uncertainty window highlighted (lower panel).}
    \label{fig:Fig_f}
\end{figure}

Dataset~A was randomly split into a training set (TS$_\mathrm{A}$) and a validation set (VS$_\mathrm{A}$), and a committee of four MLIP models was trained as described above. All committee members exhibited comparable performance. For the model selected for MD simulations, the training and validation errors at convergence remained closely aligned, with energy root means square errors (RMSEs) of 0.27 and 0.18~meV/atom and force RMSEs of 0.0259 and 0.0267~eV/\AA, respectively. The small and consistent differences between training and validation metrics indicate stable optimization and no evidence of overfitting, supporting the use of the selected model for subsequent MD simulations within the query-by-committee active-learning framework. 
A total of $162290$ atomic configurations were extracted from the 360 classical MD trajectories generated during the QBC exploration phase. Applying the predefined uncertainty criterion on atomic forces ($\Delta F$) led to the selection of $12277$ configurations for subsequent single-point DFT labeling. The energy and force distributions of these newly labeled structures are shown in Figure~\ref{fig:Fig_e} (green curves), illustrating how the active-learning step extends the training dataset toward previously underrepresented energetic and force regimes.  

Figure~\ref{fig:Fig_f} provides a quantitative overview of the active-learning exploration. The upper panel shows the distribution of configurations selected within the $\Delta F$ uncertainty window as a function of incident energy and surface temperature. As mentioned, the initial AIMD dataset was restricted to two incident energies (0.1 and 0.3~eV) and two surface temperatures (100 and 300~K). As expected, the density of selected configurations increases significantly when moving away from these original conditions. In particular, higher incident energies (above 0.5~eV) and elevated surface temperatures (above 300~K) yield substantially larger numbers of configurations exhibiting elevated model uncertainty. The highest counts are obtained at relatively low incident energies combined with elevated surface temperatures. This behavior arises from two complementary effects: thermal fluctuations of the substrate at high temperature generate configurations absent from the initial dataset, and low collision energies lead to longer residence times for NO in the interaction region near the surface, thereby increasing the number of sampled frames per trajectory. Together, these factors enhance the probability of encountering configurations associated with significant model disagreement. 
The lower panel of Figure~\ref{fig:Fig_f} shows the distribution of the QBC model-deviation score $\Delta F$ (logarithmic scale) over all extracted configurations. The distribution is highly skewed: 92.2\% of the configurations have force‑deviation scores below 0.05~eV/\AA, and only 0.24\% exceeded 0.5~eV/\AA. In other words, more than 92\% of the sampled configurations fall into the lower uncertainty region ($\Delta F < 0.05$~eV/\AA), suggesting that the potential is already close to convergence. Usually, convergence is assessed by the fraction of configurations whose model deviation falls below the lower uncertainty threshold, with values approaching 99~\% generally taken as an indication of sufficient coverage (e.g., \cite{zengDevelopmentRangeCorrectedDeep2021}). In the present case, this fraction is initially lower due to the use of a deliberately strict lower threshold. It should be noted that our lower uncertainty threshold is significantly more conservative than values commonly employed in molecule–surface studies, for example, thresholds on the order of 0.2~eV/\AA~have been reported in previous work \cite{sivakumarInitioDescriptionAdsorbate2024,zhuangResolvingOddEven2022}. When adopting this reference threshold, 98.9\% of the configurations generated during the first exploration cycle are already classified as well described. To further assess convergence, a second active-learning cycle was performed following the same computational protocol using the fine-tuned models. In this case, 99.4\% of the sampled configurations fell below the 0.2~eV/\AA~threshold, indicating a substantial reduction of extrapolative regions. On this basis, the active-learning loop was terminated after a single refinement cycle.

\begin{figure}[h]
    \centering
    \includegraphics[width=\linewidth]{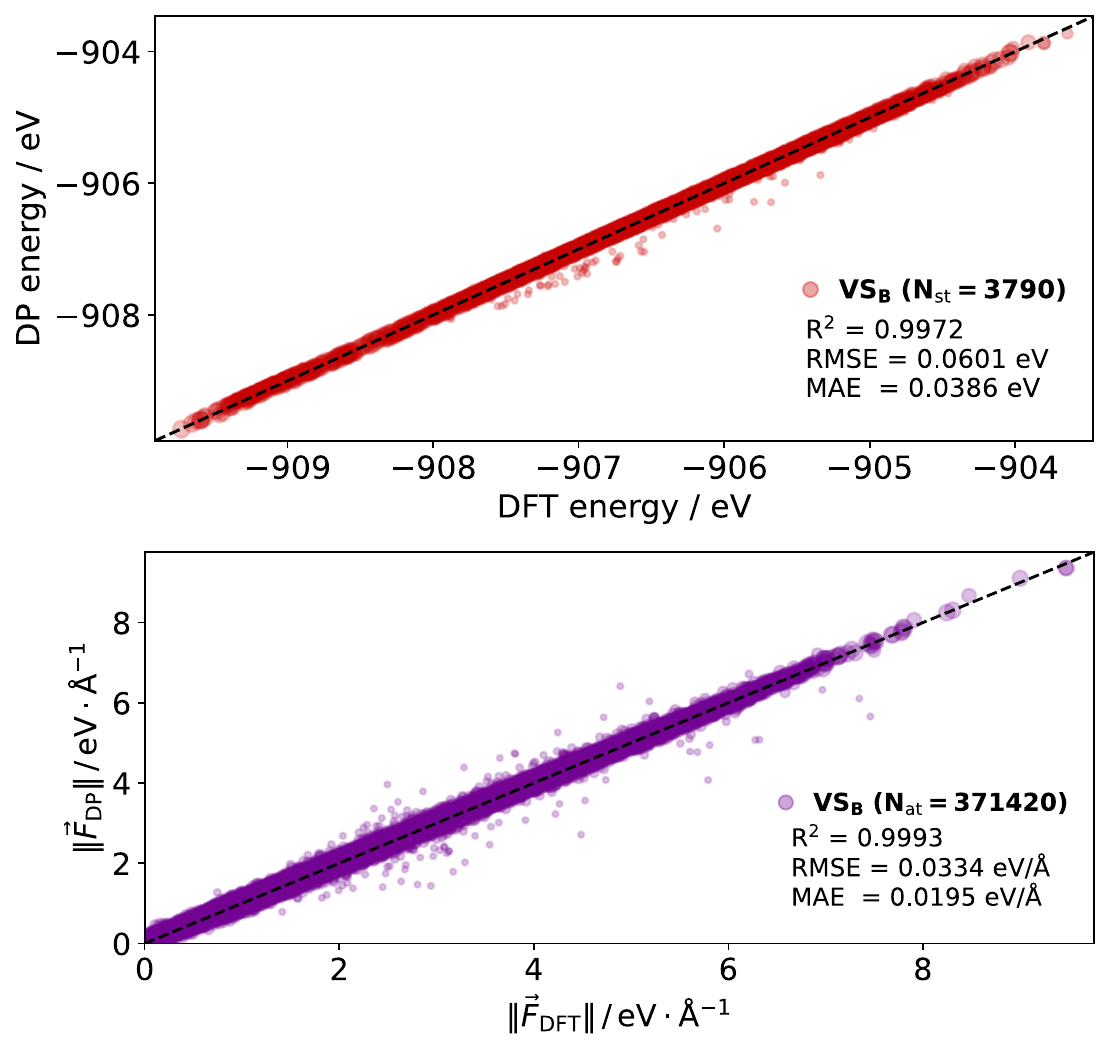}
    \caption{\textcolor{black}{Parity plots between MLIP predictions of energies (upper panel) and forces (lower panel) and the corresponding DFT reference data for validation set B)}}
    \label{fig:Fig_a}
\end{figure}

Then, the final dataset hereafter referred to as dataset~B, was constructed by combining the initial FPS-sampled dataset~A (6671 configurations) with the $12277$ configurations selected during the AL phase, resulting in a total of $18948$ labeled structures. The corresponding energy and force distributions are shown in Figure~\ref{fig:Fig_e} (orange curves), demonstrating the extended coverage of the PES achieved after active-learning refinement. 
Dataset~B was subsequently randomly split into 80\% for training (TS$_\mathrm{B}$, $N_{\text{st}} = 15159$) and 20\% for validation (VS$_\mathrm{B}$, $N_{\text{st}} = 3790$), and models were fine-tuned as described above.

To evaluate the model predictive performance on unseen data, correlation plots between DP predictions and DFT reference data for validation set~B are shown in Figure~\ref{fig:Fig_a}. The correlation on the total energy (upper panel) exhibits a $R^2$ of 0.9972, with an RMSE of 0.0601~eV and a mean square error (MAE) of 0.0386~eV. The data points lie closely along the parity line over the full energy range, indicating that the model retains excellent predictive capability across the extended configurational space.
A similarly high level of agreement is observed for atomic forces. The force correlation yields $R^2 = 0.9993$, with an RMSE of 0.0334~eV/\AA\ and a MAE of 0.0195~eV/\AA. The force predictions (lower panel) remain accurate over several orders of magnitude, including both low-force near-equilibrium configurations and high-force, strongly interacting geometries. Together, these metrics confirm that the final model trained on dataset~B achieves high fidelity with respect to DFT while maintaining robustness across the broader configurational space sampled during the QBC exploration.

\subsection{MD simulations of NO scattering on graphite}
\label{subsec:MD_sims}

\subsubsection{Adsorption energy}

Before analyzing the scattering dynamics, the adsorption energy of NO on the graphite surface was examined (computational details in Supporting Information). The most stable adsorption configuration corresponds to an adsorption energy of 142 meV, with the NO center of mass located 3.04 Å above the graphite surface. In this configuration, the molecule lies nearly parallel to the surface, with a slight tilt such that the N atom points toward the surface, and its center of mass is located approximately above the center of a graphite hexagonal ring (see Supporting Information for the geometry), in agreement with DFT-based adsorption geometries of NO on graphene \cite{houAdsorptionOxidationNO2015,chenSidopedGrapheneIdeal2012}. The adsorption energy value is consistent with recently reported calculations based on improved Lennard–Jones (ILJ) interaction potentials describing the NO–carbon interaction, which predict an adsorption energy of approximately 120~meV for a similar configuration, with the NO molecule lying nearly parallel to the surface above the center of a graphite hexagon \cite{rutiglianoScatteringNOMolecules2025}.
The present value is also in good agreement with experimental measurements, which report heats of adsorption for NO on graphitized carbon in the range 120–190 meV. \cite{brownPhysicalAdsorptionNitric1973}

\subsubsection{Scattering probabilities}

\begin{figure}[h]
    \centering
    \includegraphics[width=\linewidth]{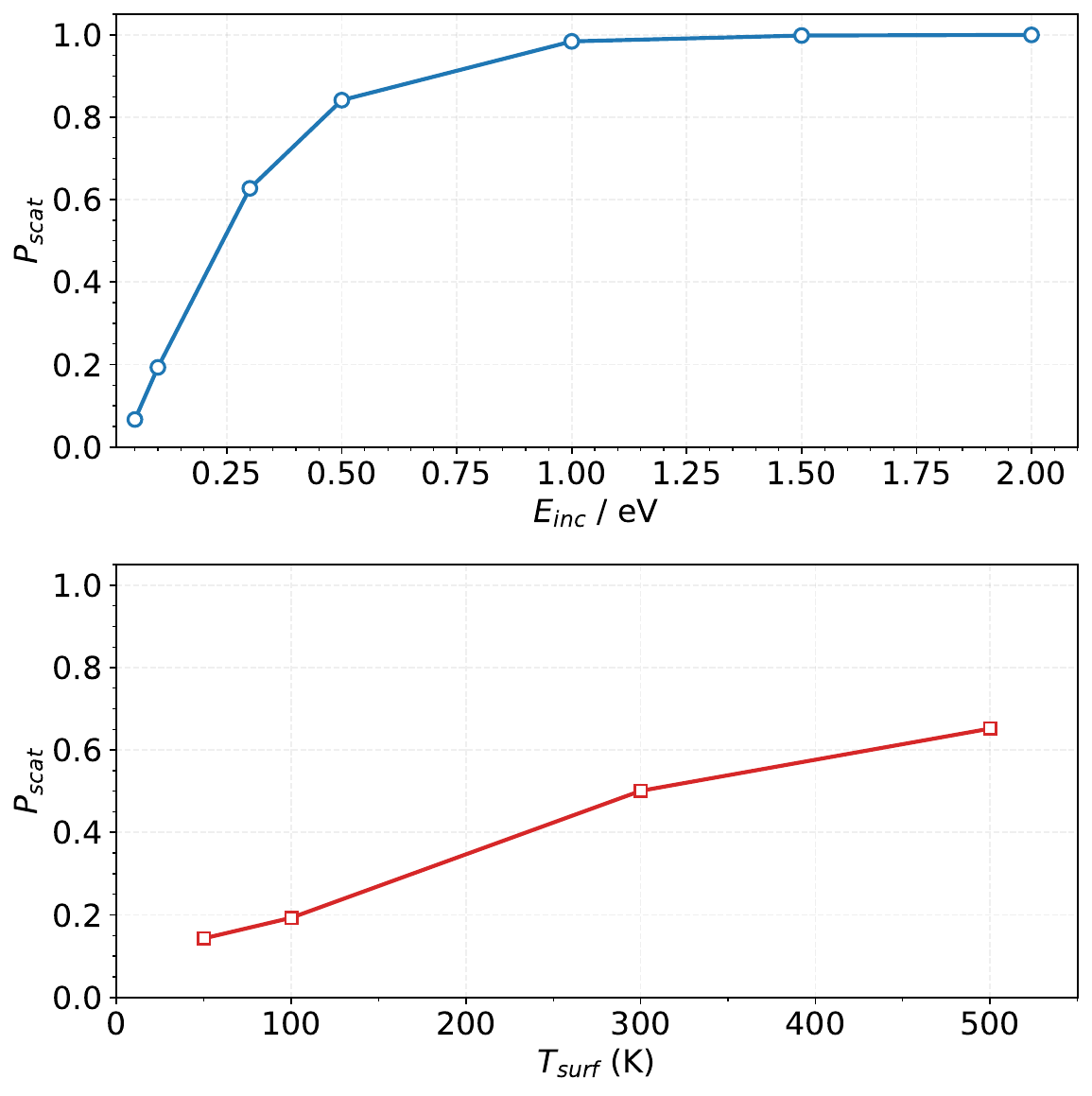}
    \caption{\textcolor{black}{Scattering probability $P_{\mathrm{scat}}$ obtained from molecular dynamics simulations. Top: dependence on the incident energy $E_{\mathrm{inc}}$ for $T_{\mathrm{surf}} = 100$~K. Bottom: dependence on the surface temperature $T_{\mathrm{surf}}$ for $E_{\mathrm{inc}} = 0.1$~eV.}} 
    \label{fig:Fig_h}
\end{figure}

\textcolor{black}{Figure \ref{fig:Fig_h} summarizes the scattering probability of NO molecules impinging on the HOPG surface obtained from our MD simulations. 
 The top panel shows the scattering probability $P_{\mathrm{scat}}$ as a function of the collision energy $E_{\mathrm{inc}}$ at a surface temperature of $T_{\mathrm{surf}} = 100$~K, while the bottom panel presents the scattering probability for a fixed collision energy of $E_{\mathrm{inc}} = 0.1$~eV as a function of the surface temperature.}

The scattering probability strongly increases with increasing collision energy, a trend that has also been observed in recent theoretical studies \cite{rutiglianoScatteringNOMolecules2025}. For the lowest collision energy ($E_{\mathrm{inc}} = 0.05$~eV), the scattering probability is very small \textcolor{black}{(6.7\%)}, indicating that most trajectories correspond to trapping-mediated events rather than direct reflection. 
This behavior is consistent with the relatively shallow physisorption well obtained in this work for NO on HOPG ($\sim0.14$~eV). Because this value corresponds to the optimized adsorption configuration at 0~K, the effective interaction energy sampled during scattering at $T_{\mathrm{surf}}=100$~K is expected to be somewhat smaller due to thermal surface fluctuations. Consequently, for collision energies below this scale the incoming molecules cannot readily escape the attractive region of the potential on their own and are predominantly transiently trapped at the surface.
However, even for collision energies above the well depth, trapping can still occur due to energy transfer during the collision. Indeed, both our simulations and previous experimental and theoretical studies \textcolor{black}{\cite{hagerScatteringRotationallyExcited1992,
meyerSurfaceScatteringDynamics2022,greenwoodNitricOxideScattering2021, greenwoodMolecularDynamicsSimulations2022}} show that NO molecules can lose a significant fraction of their translational energy upon collision with graphitic surfaces, in some cases exceeding $60\%$ of their initial translational energy. As a result, molecules with initial collision energies larger than the physisorption well depth may still become temporarily trapped if sufficient translational energy is dissipated during the collision. This energy loss mechanism and its impact on the scattering dynamics will be discussed in more detail below.

\textcolor{black}{As the collision energy increases, the probability of scattering rises rapidly and approaches unity for collision energies around $E_{\mathrm{col}} \sim 1$~eV. In this regime, trapping becomes negligible and the dynamics are dominated by direct scattering events. The overall trend therefore reflects a gradual transition from trapping-mediated dynamics at low collision energies to direct scattering at higher energies.} 

\textcolor{black}{While the top panel highlights the strong dependence of the scattering probability on the collision energy, the bottom panel of Fig.~\ref{fig:Fig_h} illustrates the influence of the surface temperature ranging from 50 to 500~K for a fixed collision energy of 0.1~eV. A clear increase in the scattering probability from about 0.1 to about 0.6 is observed as the surface temperature increases. This behavior indicates that thermal fluctuations of the surface atoms effectively increase the instantaneous corrugation of the surface potential experienced by the incoming molecule. As a consequence, the probability that the molecule leaves the surface, either by direct scattering or after transient trapping followed by desorption, increases.}
This temperature dependence is consistent with experimental studies of NO interactions with pyrolitic graphite and graphitized Pt(111) surfaces, where the sticking coefficient has been systematically measured in molecular beam experiments \cite{segnerROTATIONALSTATEPOPULATIONS}. These measurements indicate that the corresponding scattering probability increases markedly with increasing surface temperature for molecules impinging on the surface with incident kinetic energies of 80 and 210~meV at an incidence angle of 60$^\circ$ with respect to the surface normal, corresponding to perpendicular kinetic energy components $E_{\mathrm{kin}}^\perp$ of 0.02 and 0.05~eV, respectively. For example, at a normal incidence kinetic energy of about 0.02~eV, the scattering probability increases from roughly 0.3 at $T_{\mathrm{surf}} \sim 150$~K to about 0.85 at $T_{\mathrm{surf}} \sim 400$~K and approaches unity above $\sim 500$~K. A similar trend is observed at higher normal energies ($E_{\mathrm{kin}}^\perp \approx 0.05$~eV), where the scattering probability rises from approximately 0.5–0.7 near 150~K to about 0.8–0.9 at 350~K. It should be noted that the perpendicular component of the kinetic energy is particularly relevant in gas--surface scattering, as it dominates the energy transfer between the molecule and the surface. Previous studies have shown that this quantity plays a key role in determining the scattering dynamics of NO on graphite \cite{hagerScatteringNOGraphite2004}. 

\subsubsection{Speed distributions and kinetic energy loss of scattered molecules}

\begin{figure}[h]
    \centering
    \includegraphics[width=0.98\linewidth]{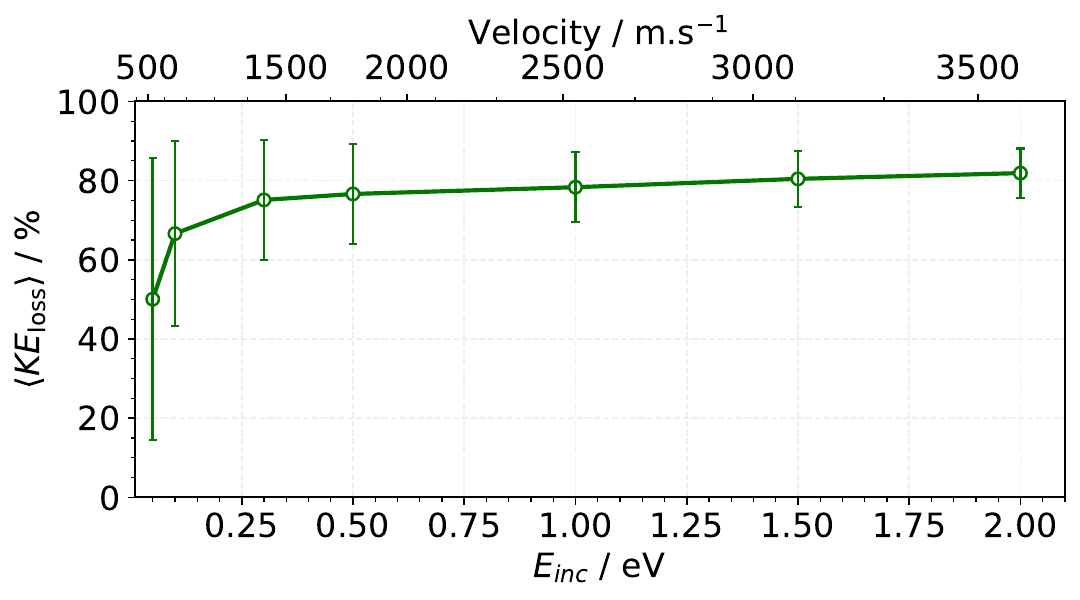}
    \includegraphics[width=\linewidth]{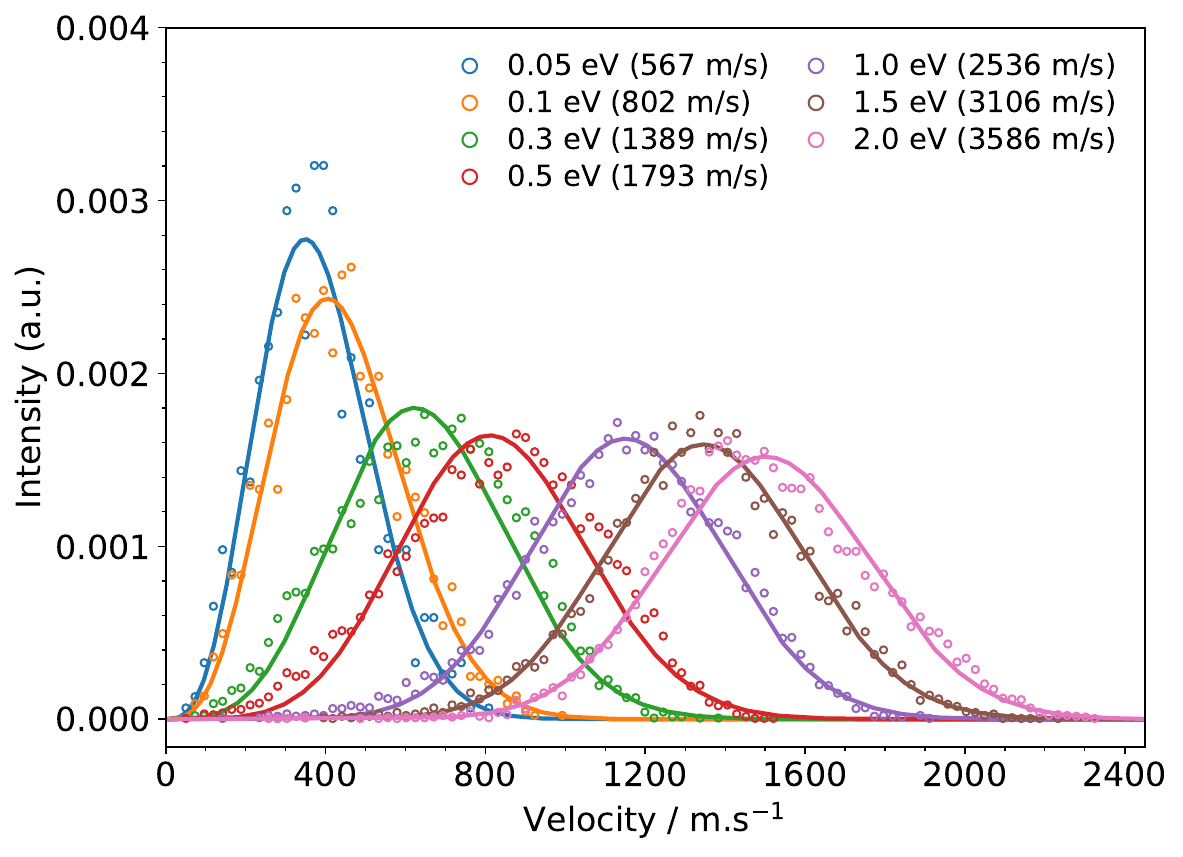}
    \caption{\textcolor{black}{Top: Mean kinetic energy loss of scattered NO as a function of incident energy $E_{\mathrm{inc}}$. Error bars denote one standard deviation. Bottom: Corresponding speed distributions of the scattered molecules. Solid lines show fits to a three-dimensional flux distribution \cite{greenwoodMolecularDynamicsSimulations2022}}. The analytical fitted function and its parameters are given in table \ref{tab:flux_fit_parameters}. All data correspond to scattering from graphite at $T_{\mathrm{surf}}$ = 100 K for incident energies between 0.05 and 2.0 eV.}
    \label{fig:Fig_g}
\end{figure}

\begin{table}[h]
\centering
\caption{Parameters obtained from the three-dimensional flux distribution fits of the scattered NO velocity distributions at $T_{\mathrm{surf}}=100$~K for different incident energies $E_{\mathrm{inc}}$. The fitted function is $F(c)=A\,c^{3}\exp[-(c-c_0)^2/\alpha^2]$ \cite{greenwoodMolecularDynamicsSimulations2022}. $v_{\text{peak}}$ corresponds to the velocity at which the flux distribution reaches its maximum, i.e., the most probable speed of the scattered molecules in the flux-weighted velocity distribution. \\}
\label{tab:flux_fit_parameters}
\setlength{\tabcolsep}{3pt}
\small
\begin{tabular}{c c c c c}
\hline
$E_{\mathrm{inc}}$ (eV) & $A$ & $c_0$ & $\alpha$ & $v_{\mathrm{peak}}$ \\
 &  & (m s$^{-1}$) & (m.s$^{-1}$) & (m.s$^{-1}$) \\
\hline
0.05 & $2.9\!\times\!10^{-10}$ & 1.7 & 284.5 & 349.4 \\
0.10 & $1.6\!\times\!10^{-10}$ & 3.2 & 329.5 & 405.1 \\
0.30 & $1.7\!\times\!10^{-11}$ & 275.2 & 380.1 & 623.0 \\
0.50 & $5.1\!\times\!10^{-12}$ & 543.2 & 380.4 & 810.8 \\
1.00 & $1.3\!\times\!10^{-12}$ & 977.1 & 365.1 & 1150.8 \\
1.50 & $7.7\!\times\!10^{-13}$ & 1194.3 & 369.6 & 1346.5 \\
2.00 & $5.2\!\times\!10^{-13}$ & 1355.3 & 382.4 & 1501.3 \\
\hline
\end{tabular}
\end{table}

\textcolor{black}{Beyond the scattering probability, the MLIP developed in this work enables extensive dynamical simulations that provide statistically meaningful information on the properties of the scattered molecules. In particular, it allows us to characterize both the speed distributions of the scattered molecules and the associated translational energy loss during the scattering event, quantities that can potentially be compared with experimental measurements.}

The evolution of the translational energy transfer during the scattering process is summarized in Fig.~\ref{fig:Fig_g}. The top panel shows the mean kinetic energy loss of the scattered molecules as a function of the incident energy at $T_{\mathrm{surf}}=100$~K, with error bars corresponding to one standard deviation obtained from the ensemble of trajectories. The bottom panel displays the corresponding velocity distributions for each incident energy of the scattered NO molecules together with fits to a three-dimensional flux distribution $F(c)=A\,c^{3}\exp[-(c-c_0)^2/\alpha^2]$ \cite{greenwoodMolecularDynamicsSimulations2022}. The fitted parameters, together with the most probable speed of the scattered molecules in the flux-weighted velocity distribution ($v_{\mathrm{peak}}$), are reported in Table~\ref{tab:flux_fit_parameters}.

We observed that the scattered molecules lose between approximately 50\% and 82\% of their initial kinetic energy over the range of incident energies investigated. The mean kinetic energy loss increases rapidly from about 50\% at $E_{\mathrm{inc}}=0.05$~eV to roughly 75\% at $E_{\mathrm{inc}}=0.3$~eV, and then increases more gradually to about 82\% at $E_{\mathrm{inc}}=2.0$~eV. At the same time, the standard deviation of the kinetic energy loss decreases with increasing incident energy. This trend indicates that the dispersion of collision outcomes becomes smaller at higher incident energies, reflecting a transition toward a more impulsive and deterministic scattering regime where the fraction of translational energy transferred to the surface becomes more narrowly distributed. At fixed incident energy ($E_{\mathrm{inc}}=0.1$~eV), the mean kinetic energy loss decreases with increasing surface temperature, from about 72\% at 50~K to approximately 28\% at 500~K, while the dispersion of the distribution increases (from $\pm 19\%$ to $\pm 53\%$; see Supporting Information). This behavior indicates that higher surface temperatures promote more efficient energy return to the molecule through thermal motion of surface atoms, leading to less net energy transfer to the surface and a broader range of scattering outcomes.
Overall, the magnitude of the kinetic energy loss is consistent with previous experimental studies, with NO scattering from graphene reporting losses of about 80\% at $E_{\mathrm{inc}}=0.31$~eV \cite{greenwoodNitricOxideScattering2021}. Losses of around 66\% are also mentioned in \cite{meyerSurfaceScatteringDynamics2022} for NO–graphite. 

The analysis of the velocity distributions provides further insight into the energy transfer mechanisms during the scattering process.
At the two lowest incident energies ($E_{\mathrm{inc}} = 0.05$ and $0.10$~eV), the fitted distributions are characterized by center parameters $c_0$ close to zero (1.7 and 3.2~m~s$^{-1}$, respectively), indicating that the Gaussian envelope is essentially centered at zero velocity, meaning that the intrinsic velocity distribution is not shifted toward the incident velocity. In other words, the scattered molecules largely lose memory of their incident velocity as a result of strong energy exchange with the surface. In this case the overall shape of the flux distribution is largely determined by the $c^3$ pre-factor. This behavior indicates that the dynamics are strongly influenced by the attractive part of the molecule–surface interaction potential and is consistent with a mechanism in which a significant fraction of the molecules are captured transiently in the adsorption well and partially thermally accommodated to the surface before desorption. Interestingly, at these two lowest incident energies, a fraction of the scattered molecules exhibit exit velocities exceeding the incident beam velocity. This behavior is visible in the velocity distributions, where the high-velocity tail of the flux distribution extends beyond the incoming speed. 
Such events indicate that some molecules acquire additional kinetic energy from the thermal motion of the surface prior to desorption. However, this feature progressively vanishes with increasing collision energy and is no longer observed for incident energies above 0.3 eV.
A marked transition in the velocity distributions occurs between $E_{\mathrm{inc}} = 0.10$ and $0.30$~eV, where the fitted parameter $c_0$ increases significantly (to $275$~m~s$^{-1}$) and the most probable velocity shifts to $v_{\mathrm{peak}} = 623$~m~s$^{-1}$. This evolution indicates that the scattered molecules retain a larger fraction of their initial translational momentum, signaling a transition from a regime strongly influenced by transient trapping and partial thermal accommodation within the adsorption well to a regime dominated by direct molecule–surface collisions. Then, 
for $E_{\mathrm{inc}} \geq 0.30$~eV, both $c_0$ and $v_{\mathrm{peak}}$ increase steadily with the incident energy, while the velocity distributions retain a similar overall shape, as reflected by the nearly constant width parameter $\alpha$ ($\alpha \approx 365$–382~m~s$^{-1}$). This indicates that the same scattering mechanism operates across this energy range. The nearly constant fractional kinetic energy loss ($\sim75$–80\%) further suggests that a fixed proportion of the incident translational energy is transferred to the surface during the collision. As a result, the outgoing velocities scale primarily with the incoming velocity, consistent with a direct impulsive scattering regime in which the molecule undergoes one or a few strong momentum-exchange events with the surface atoms.

\subsubsection{Polar angle distributions}

\begin{figure}[h]
    \centering
    \includegraphics[width=\linewidth]{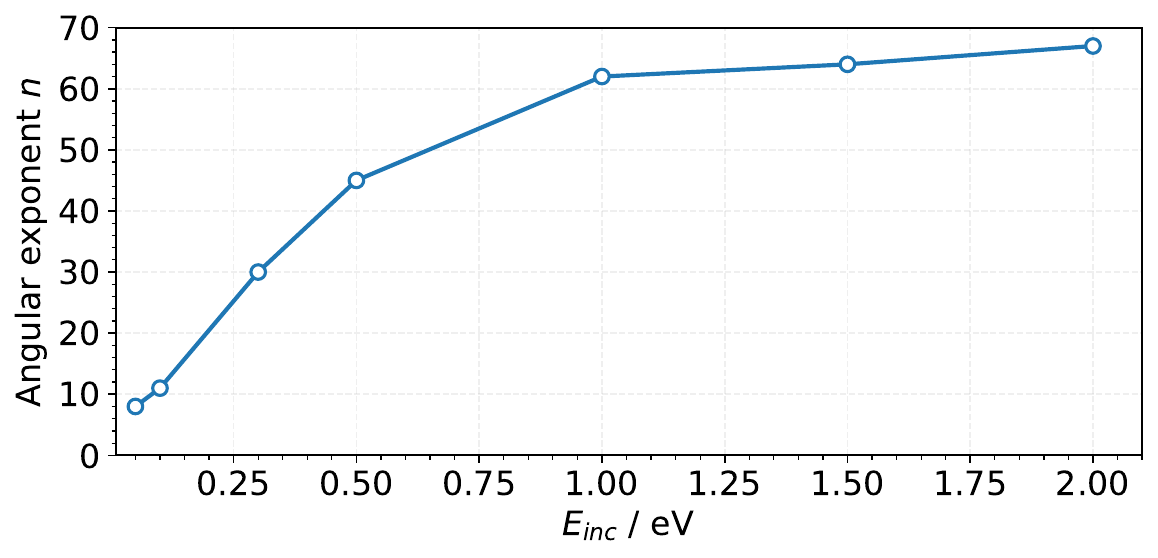}
    \caption{\textcolor{black}{Evolution of the angular scattering exponent $n$ as a function of the incident translational energy $E_{\mathrm{inc}}$ for $T_{\mathrm{surf}}$ = 100 K. The exponent $n$ is obtained by fitting the polar angular distribution of scattered trajectories with a $\cos^{n}(\theta_{\mathrm{scatt}})$ function.}}
    \label{fig:Fig_i}
\end{figure}

\textcolor{black}{Complementary insight into the scattering dynamics can be obtained from the polar angular distributions of the scattered molecules, which provide information on the scattering mechanism and the corrugation of the gas–surface interaction potential. In particular, these distributions characterize the preferred scattering directions of the outgoing molecules and constitute key observables for interpreting molecular beam scattering experiments.}
The polar angular distributions of the scattered molecules were obtained by binning the outgoing scattering angles $\theta_{\mathrm{scatt}}$ (angle between the surface normal and the molecule center-of-mass) in 5° intervals between 0° and 90°. To account for the geometrical dependence of the solid angle, the histograms were corrected by dividing the counts by $\sin\theta$, yielding an intensity proportional to the probability per unit solid angle $dP/d\Omega$. For each incident energy, the corrected distributions were normalized to their maximum value and fitted with a $\cos^n(\theta_{\mathrm{scatt}})$ function \cite{greenwoodNitricOxideScattering2021}. 
Both raw and solid-angle–corrected distributions are provided in the Supporting Information.
The angular component $n$ is expected to be equal to unity for pure thermal desorption \cite{comsaDynamicalParametersDesorbing1985}. Increasing values of $n$ indicate a progressively narrower angular distribution centered around the surface normal. The evolution of $n$ with $E_{\mathrm{inc}}$ and $T_{\mathrm{surf}}$ = 100~K is shown on Figure \ref{fig:Fig_i}. 
The fitted angular exponents show a strong dependence on the incident energy. In the low-to-intermediate energy range ($E_{\mathrm{inc}}$ = 0.05–0.5 eV), the exponent increases rapidly from $n$ = 8 to $n$ = 45, indicating a progressive narrowing of the angular distribution and a transition toward more forward-focused scattering. 
At higher incident energies, the exponent reaches a plateau with $n$~=~62~–~67 for $E_{\mathrm{inc}} \ge 1.0$~eV. This saturation shows that the angular distributions have approached a strongly forward-peaked regime in which further increases in collision energy no longer significantly modify the scattering directionality. In this regime, the dynamics are dominated by short interaction times with the surface, resulting in nearly specular reflection of the incident momentum component normal to the surface. 
In our simulations, the surface temperature does not significantly influence the angular distributions of the scattered molecules. Similar distributions, characterized by comparable $n$~=~10-11 values, are obtained for all four surface temperatures considered at an incident energy of $E_{\mathrm{inc}} = 0.1$~eV (see Supporting Information). 
The evolution of the angular distributions with incident kinetic energy for the NO–graphite and graphene systems is well established both experimentally and theoretically. Molecular beam studies and trajectory calculations show that increasing the incident kinetic energy progressively focuses the scattered molecules toward the specular direction \cite{segnerROTATIONALSTATEPOPULATIONS,hagerLaserInvestigationDynamics1985,greenwoodMolecularDynamicsSimulations2022}. This behavior is well reproduced in the present simulations.
In contrast, experimental studies indicate that the surface temperature primarily affects the relative contributions of the diffuse and specular components of the angular distributions. In particular, the diffuse scattering contribution decreases with increasing surface temperature and becomes negligible above $T_{\mathrm{surf}} \sim 500$~K \cite{frenkelRotationalStatePopulations1982,meyerSurfaceScatteringDynamics2022,hagerScatteringNOGraphite2004}, while a slight rotation of the specular lobe toward the specular direction is observed \cite{hagerLaserInvestigationDynamics1985,hagerRotationallyExcitedNO1997}. Such effects are not clearly observed in the present simulations, likely due to the limited simulation time, which restricts the sampling of trapping–desorption events responsible for the diffuse component, as well as to the normal-incidence collision conditions considered here.
Indeed, the residence time of NO on graphite has been estimated to be on the order of $\sim10$~ps at $T_{\mathrm{surf}}=300$~K \cite{vachSurvivalRelaxationExcitation1989}, increasing up to $\sim10^{-9}$~s at $T_{\mathrm{surf}}=150$~K and decreasing to $\sim1$~ps at $T_{\mathrm{surf}}=700$~K \cite{frenkelRotationalStatePopulations1982,meyerSurfaceScatteringDynamics2022}. These time scales are comparable to or significantly longer than the maximum trajectory duration considered in the present simulations, which likely limits the observation of trapping-mediated diffuse scattering events.

\subsubsection{Ro-vibrational states of scattered NO}

\textcolor{black}{The state-resolved observables constitute particularly sensitive probes of the collision dynamics, as they reflect how energy is redistributed among the molecular degrees of freedom during the interaction with the surface. The ro-vibrational state distributions of the scattered NO molecules therefore provide further insight into the NO–HOPG interaction dynamics. As discussed above, the use of the MLIP enables the generation of large ensembles of scattering trajectories, ensuring robust statistics for the ro-vibrational populations obtained from the simulations. Such information offers detailed insight into the microscopic mechanisms governing gas–surface interactions and the energy transfer pathways between translational, rotational, and vibrational modes, and is closely related to observables measured in molecular beam scattering experiments. In the following, we analyze the ro-vibrational states of the scattered NO molecules.}
As mentioned above, the rovibrational state of scattered NO was determined using semiclassical quantization. For each incident energy, an effective rotational temperature $T_\mathrm{rot}$ was extracted from the ensemble of scattered trajectories by maximum-likelihood fitting of the rotational population distribution to a Maxwell-Boltzmann (MB) model assuming rigid-rotor level populations $p_j \propto (2j+1)\exp[-E_j/(k_\mathrm{B}T_\mathrm{rot})]$. The reported uncertainty in $T_\mathrm{rot}$ corresponds to the statistical uncertainty of the fitted parameter, estimated from the curvature of the log-likelihood function at its maximum. 
Consistent with previous experimental and theoretical studies \cite{rutiglianoScatteringNOMolecules2025,vachEnergyTransferProcesses1987,  greenwoodVelocitySelectedRotationalState2023, meyerSurfaceScatteringDynamics2022}, no vibrational excitation from $v = 0$ to higher vibrational states was observed in any of the simulations. The NO molecule therefore retains its initially assigned vibrational energy throughout the trajectories, even at the highest incident energies and surface temperatures considered in this work.

\paragraph{Rotational Temperature as a Function of Incident energy}

\begin{figure*}[h!]
    \centering
    \includegraphics[width=\linewidth]{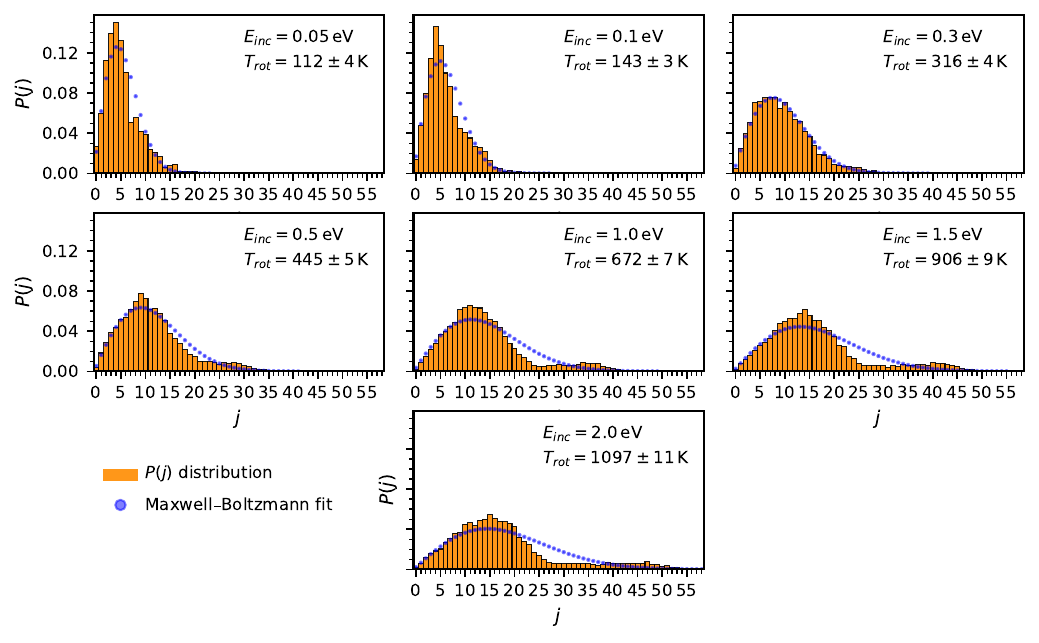}
    \caption{\textcolor{black}{Rotational state distributions of scattered NO for incident energies between 0.05 and 2.0~eV at a surface temperature of 100~K. Maxwell–Boltzmann fits are shown as discrete markers superimposed on the distributions.}}
    \label{Fig_j}
\end{figure*}

The rotational state distributions of scattered NO  from graphite at $T_{\mathrm{surf}} = 100$~K for different incident energies  are shown in Figure~\ref{Fig_j}. Two distinct regimes can be identified depending the  incident energies considered. For $E_{\mathrm{inc}} \le 0.3$~eV, the rotational distributions are  well described by a MB model. At the lowest collision energies ($E_{\mathrm{inc}} = 0.05$ and $0.1$~eV), the distributions are strongly peaked at low rotational quantum numbers ($j \approx 3$-6) and decay rapidly toward higher $j$ values, corresponding to effective rotational temperatures of $T_{\mathrm{rot}} \approx 112$~K and $143$~K, respectively. These values are consistent with the experimental results of Hager \textit{et al.} \cite{hagerScatteringNOGraphite2004}, who reported a specular rotational temperature of approximately $T_\mathrm{rot} \sim 150$~K for NO scattering from graphite at $T_{\mathrm{surf}} \approx 100$~K and a normal component of the kinetic energy of $E_{\mathrm{kin}}^\perp \approx 0.1$~eV.
At $E_{\mathrm{inc}} = 0.3$~eV the distribution becomes significantly broader and the maximum shifts toward higher rotational states ($j \approx 8$). The corresponding rotational temperature increases to $T_{\mathrm{rot}} \approx 316$~K, reflecting a larger transfer of translational energy into rotational motion during the collision.  
Deviations from the MB model become significant for $E_{\mathrm{inc}} \ge 0.5$~eV, where the rotational distributions develop pronounced high-$j$ tails. In this second regime, two distinct features can be identified. The first component is a bell-shaped distribution that progressively broadens and whose maximum shifts toward higher rotational states, from $j \approx 8$-10 at $E_{\mathrm{inc}} = 0.5$~eV to $j \approx 14$-15 at $E_{\mathrm{inc}} = 2.0$~eV. This part of the distribution corresponds to the dominant population of direct scattering trajectories, for which the molecule experiences moderate torques during the collision with the surface. Because a wide range of collision geometries contributes to these events, the transferred angular momentum is distributed over many rotational states, producing a smooth bell-shaped distribution.
The second component appears as a high-$j$ tail corresponding to a minority of highly impulsive trajectories in which the molecule experiences particularly strong torques during the interaction with the surface. These events lead to extreme rotational excitation and are commonly interpreted as the signature of rotational rainbow scattering \cite{rutiglianoScatteringNOMolecules2025,nymanSurfaceScatteringNO1990,meyerSurfaceScatteringDynamics2022,hagerScatteringNOGraphite2004}. In this case, specific collision geometries maximize the torque exerted by the anisotropic NO–graphite interaction, causing a concentration of trajectories that populate very large rotational quantum numbers.

\paragraph{Rotational Temperature as a Function of Surface Temperature}

\begin{figure}[h]
    \centering
    \includegraphics[width=\linewidth]{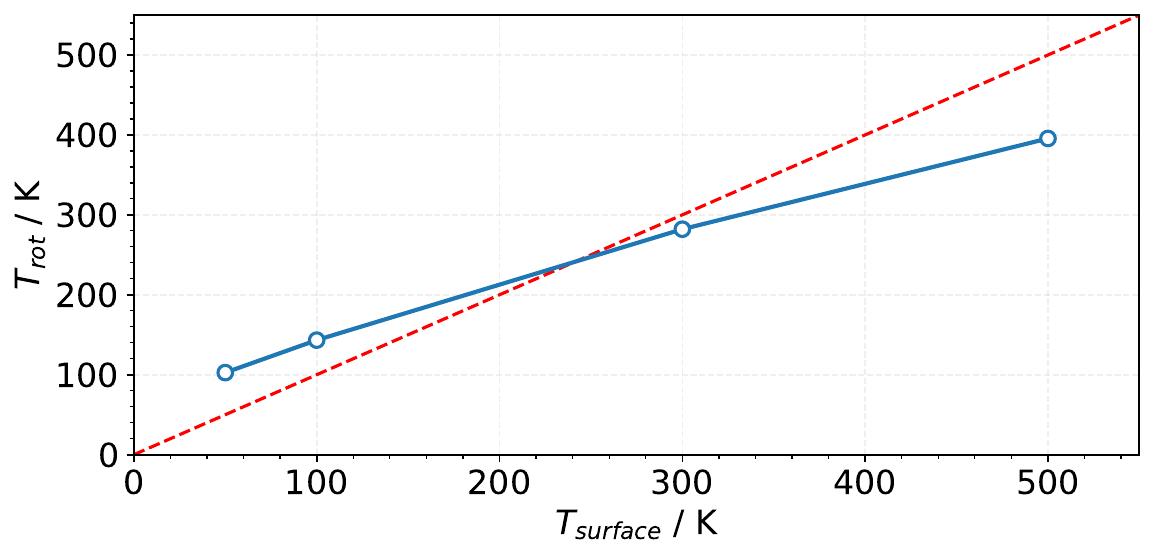}
    \caption{\textcolor{black}{Rotational temperature of scattered NO as a function of surface temperature $T_{\mathrm{surf}}$ at $E_{\mathrm{inc}}~=~0.1$~eV. The dashed line corresponds to $T_{\mathrm{rot}} = T_{\mathrm{surf}}$, indicating the limit of complete rotational accommodation.}}
    \label{fig:Fig_k}
\end{figure}

The evolution of the rotational temperature of scattered NO for $E_{\mathrm{inc}}=0.1$~eV at different surface temperatures $T_{\mathrm{surf}}$ is shown in Figure~\ref{fig:Fig_k}. For all investigated temperatures, the rotational state distributions are well described by a MB distribution (see Supporting Information), allowing the assignment of an effective rotational temperature $T_{\mathrm{rot}}$. Experimentally, NO scattering from graphite exhibits a well-known temperature dependence characterized by an intermediate regime of rotational accommodation ($T_{\mathrm{rot}} \approx T_{\mathrm{surf}}$), followed by saturation of $T_{\mathrm{rot}}$ at both high and low surface temperatures \cite{frenkelRotationalStatePopulations1982,vachEnergyTransferProcesses1987,hagerScatteringNOGraphite2004}. Furthermore, it has been shown that increasing the incident kinetic energy shifts the absolute value of $T_{\mathrm{rot}}$ upward while preserving this saturation behavior \cite{hagerScatteringNOGraphite2004}.
In the present simulations, $T_{\mathrm{rot}}$ increases monotonically with the considered surface temperatures but does not strictly follow the line of complete rotational accommodation ($T_{\mathrm{rot}} = T_{\mathrm{surf}}$). At low surface temperatures ($T_{\mathrm{surf}}=50$–100~K), the extracted rotational temperatures exceed the surface temperature ($T_{\mathrm{rot}}>T_{\mathrm{surf}}$), indicating as experiments that rotational excitation is primarily driven by the conversion of incident translational energy during the collision. Around $T_{\mathrm{surf}}\approx200$–300~K, $T_{\mathrm{rot}}$ approaches the accommodation limit, suggesting efficient energy exchange between the molecular rotation and surface phonons. This near-accommodation behavior around $T_{\mathrm{surf}} \sim 250$~K is quantitatively consistent with experimental observations for NO scattering from graphite at a normal component of the incident kinetic energy of $E_{\mathrm{kin}}^\perp \approx 0.1$~eV \cite{hagerScatteringNOGraphite2004}.
At higher temperatures ($T_{\mathrm{surf}}=500$~K), $T_{\mathrm{rot}}$ remains significantly below the accommodation line. Such sub-thermal rotational temperatures at high $T_{\mathrm{surf}}$ are consistent with aforementioned experimental observations for NO scattering from graphite, where $T_{\mathrm{rot}}$ saturates at values well below the surface temperature, a behavior attributed to the near-conservation of the angular momentum component normal to the atomically flat graphite surface \cite{nymanSurfaceScatteringNO1990}. 

\section{Conclusion}
\label{sec:ccl}

\textcolor{black}{We have presented a data driven strategy for the construction of a machine learning interatomic potential tailored to gas surface scattering, using NO on HOPG as a representative benchmark system. Starting from AIMD trajectories, the workflow combines SOAP based structural descriptors, principal component analysis, and farthest point sampling to identify a compact set of configurations that captures the essential configurational diversity of the NO--HOPG interaction. This analysis made it possible to identify the most relevant regions of configurational space for the scattering process and to construct an initial training dataset that is both compact and physically meaningful.}

\textcolor{black}{On this basis, Deep Potential models were trained and subsequently refined through a query-by-committee active learning protocol. The active learning stage proved particularly efficient, since only a single refinement cycle was required to obtain a robust final MLIP with excellent agreement with the DFT reference energies and forces over the extended configurational space explored during the simulations. The resulting potential thus combines near \textit{ab initio} accuracy with the computational efficiency needed for extensive molecular dynamics simulations.}

\textcolor{black}{The final MLIP was then used in \textsc{LAMMPS} to investigate NO scattering from HOPG over broad ranges of collision energy and surface temperature through large trajectory ensembles (over $10^5$ trajectories) that would be prohibitively expensive at the AIMD level. These simulations provide a detailed and statistically robust atomistic description of the NO--HOPG interaction dynamics, including the balance between trapping and direct scattering, the translational energy loss, the evolution of polar angular distributions, and the rotational excitation of the scattered molecules. Overall, the calculated observables reproduce the main trends reported experimentally and computationally, and provide a consistent microscopic interpretation of the underlying energy transfer mechanisms. The NO--HOPG scattering dynamics are characterized by: (i) a transition from trapping-mediated dynamics at low incident energies ($E_{\mathrm{inc}} \leq 0.1$~eV), characterized by low scattering probabilities ($<20\%$), large energy dissipation (50--70\%), and loss of velocity memory, to a direct impulsive scattering regime at higher energies where the scattering probability approaches unity; (ii) a nearly constant fractional translational energy loss of $\sim75$--80\% for $E_{\mathrm{inc}} \geq 0.3$~eV, leading to outgoing velocities that scale with the incident velocity; and (iii) a progressive narrowing and forward focusing of the angular distributions, together with increasing rotational excitation and the emergence of high-$j$ tails associated with rotational rainbow scattering at high energies, while the rotational temperature increases with surface temperature but exceeds $T_{\mathrm{surf}}$ at low temperatures, reflecting dominant translation-to-rotation energy transfer, and remains sub-thermal at high $T_{\mathrm{surf}}$, indicating incomplete rotational accommodation. In contrast, no vibrational excitation is observed over the range of collision energies and surface temperatures investigated, indicating that the scattering remains vibrationally elastic under the present conditions.}

\textcolor{black}{More generally, this work demonstrates that the combination of descriptor-guided sampling, dimensionality reduction, and query-by-committee active learning provides an efficient and transferable route for the development of MLIPs for gas--surface interactions. Beyond the specific NO--HOPG case, the present strategy provides a solid basis for future applications to more complex and potentially reactive molecule--surface systems, where broad configurational coverage of reaction pathways, robust statistics, and near \textit{ab initio} accuracy are required. In this perspective, the present work highlights how data-driven potential construction combined with active learning can bridge the gap between first-principles accuracy and the statistical requirements of gas--surface scattering simulations. This framework is naturally compatible with the emerging generation of foundation models for atomistic simulations, which can be leveraged as transferable initial representations and subsequently refined for specific gas--surface systems through targeted active learning.}

\section*{Acknowledgements}

This work was funded by the CNRS Émergence@Physique 2025 project. The authors acknowledge support from the French national supercomputing facilities (Grant Nos. DARI A0190801859, A0170801859) and from the Centre de Ressources Informatiques (CRI) of the Université de Lille.

\section*{Abbreviations}

AIMD, \textit{ab initio} molecular dynamics; \\
MLIP, machine-learning interatomic potential; \\
DFT, density functional theory; \\
PES, potential energy surface; \\
PAW, projector augmented-wave; \\
VASP, Vienna \textit{Ab initio} Simulation Package \\
PBE, Perdew--Burke--Ernzerhof \\
MD, molecular dynamics; \\
LAMMPS, Large-scale Atomic/Molecular Massively Parallel Simulator \\
QCT, quasi-classical trajectory; \\
SOAP, smooth overlap of atomic positions; \\
PCA, principal component analysis; \\
FPS, farthest point sampling; \\
QBC, query-by-committee; \\
DP, Deep Potential; \\
HOPG, highly oriented pyrolytic graphite; \\
RMSE, root mean square error; \\
MAE, mean absolute error.\\

\section*{Keywords}

Gas--surface scattering; \\ 
Machine-learning interatomic potentials; \\ 
Active learning; \\ 
Deep Potential; \\ 
Graphite; \\ 
Nitric oxide; \\ 
Molecular dynamics; \\ 
Surface dynamics

\section*{Supporting information}

Supporting Information contains additional computational details and complementary results on NO scattering dynamics, as well as the geometry corresponding to the most stable adsorption configuration.

\printbibliography

\end{document}